\documentclass[conference]{IEEEtran}

\usepackage{cite}
\usepackage{graphicx}
\usepackage{amsmath,amssymb}
\usepackage[T1]{fontenc}
\usepackage{url}
\usepackage{booktabs}
\usepackage{array}
\usepackage{listings}
\usepackage{xcolor}

\newcommand{\multirow}[3]{#3}
\newcommand{\localfigcaption}[1]{\caption{#1}}

\begin{document}

\title{LLM-Based Porting of Optimized C++ to CUDA Through Deoptimization and Reoptimization}

\author{\IEEEauthorblockN{Daichi Mukunoki, Ryo Mikasa, Shunichiro Hayashi,
Tetsuya Hoshino and Takahiro Katagiri}
\IEEEauthorblockA{Nagoya University\\
Furo-cho, Chikusa-ku, Nagoya 464-8601, Japan\\
mukunoki.daichi.p2@f.mail.nagoya-u.ac.jp}}

\maketitle

\begin{abstract}
When porting high-performance computing (HPC) code from CPU to GPU, CPU-oriented optimizations may obstruct LLM-based CUDA translation.
We design and evaluate a Deopt-Reopt workflow that first simplifies the input C++ code and then retranslates and reoptimizes it for CUDA, comparing it against direct translation (Direct) on twelve HPC kernels with two LLMs (\texttt{gpt-oss-120b} (O120) and \texttt{qwen-3-235b-a22b-instruct-2507} (Q235)) in Single-shot (one pass) and Iterative (repeated refinement) settings.
In Single-shot, among 18 testable cases Deopt-Reopt was significantly faster among successful trials (after BH-FDR correction) in five --- most clearly for \texttt{conv2d}, where CPU- and GPU-oriented designs diverge --- but Direct was faster in three, so removing CPU-specific optimizations is not universally beneficial.
An exploratory Direct-3 control that equalizes the LLM-call count left Deopt-Reopt ahead in only four of nineteen testable cases, with Direct-3 ahead in four others.
In Iterative, repeated generation and repair narrow the mode gap --- markedly so for O120 --- while Q235 retains large Deopt-Reopt advantages on \texttt{conv2d}, \texttt{ddgemm}, and \texttt{bgemm}.
Deopt-Reopt's effect on feasibility is also mixed --- sharply higher for some kernels Direct rarely compiles, lower for others. Because performance is conditioned on successful trials, the benefit is conditional rather than a guaranteed end-to-end gain. Overall, Deopt-Reopt is an effective but non-universal technique for LLM-based GPU porting, with gains that depend on the kernel, the model, the search budget, and the success rate.
\end{abstract}


\section{Introduction}

Scientific computing code is often developed from a simple implementation and then optimized for a specific architecture, but the simple version may be lost or never maintained.
When such CPU-oriented code is ported to a GPU, handwritten optimizations such as SIMD expansion, blocking, and data-layout transformations may obscure the algorithmic structure needed for a different parallelization strategy.
In such cases, deoptimization (removing architecture-specific optimizations before translation) may improve portability, although it may also discard structure that direct translation could have preserved.

In recent years, the code generation capability of large language models (LLMs) has improved rapidly, and their application to automatic translation and optimization of HPC code has been actively studied~\cite{dhruv2025codescribe,coderosetta2024,chen2024fortran2cpp}.
Most existing work, however, focuses on direct translation from CPU code to GPU code or on one-way optimization.
The effectiveness of such deoptimization before LLM-based GPU porting has not been sufficiently investigated.

This paper studies the effectiveness of deoptimization in LLM-based GPU porting of CPU code from C++ to CUDA.
We compare two workflows: Direct, which translates the CPU code straight to CUDA, and Deopt-Reopt, which first deoptimizes and then translates and reoptimizes.
We evaluate both in a Single-shot (one pass) and an Iterative (repeated repair and selection) setting, using two LLMs, OpenAI's \texttt{gpt-oss-120b} (O120) and Alibaba Cloud's \texttt{qwen-3-235b-a22b-instruct-2507} (Q235), on twelve HPC kernels.
The analysis focuses on when deoptimization helps, when reoptimization recovers performance, and when Iterative refinement narrows the gap between the workflows.
We use feasibility for the property that translated code compiles, runs, and passes validation, and report performance only over successful trials (successful-trial performance), separately from feasibility.

We find that deoptimization is effective but not universal: it helps most when CPU- and GPU-oriented designs diverge, yet offers little or even hurts where direct translation already preserves a usable structure, so it is best applied selectively rather than by default.
Iterative refinement narrows the gap between the workflows for some models while leaving large Deopt-Reopt advantages on a few kernels, and an exploratory call-count-matched control suggests that part of the Single-shot benefit reflects the extra LLM calls rather than deoptimization itself.

The rest of this paper is organized as follows.
Section~\ref{sec:related} reviews related work on LLM-based code translation and GPU porting.
Section~\ref{sec:experiment} describes the common evaluation setting --- the porting task, hardware and software environment, models, metrics, and benchmark kernels.
Section~\ref{sec:simple_workflow} presents the Single-shot workflow and its results, and Section~\ref{sec:iterative} the Iterative workflow.
Section~\ref{sec:conclusion} concludes.

\section{Related Work}
\label{sec:related}

CPU-to-CUDA porting has long been studied through compiler- and directive-based approaches such as OpenMP-to-GPGPU~\cite{lee2009openmp} and PPCG~\cite{verdoolaege2013ppcg}, which rely on analyzable dependences or annotations rather than LLM-based translation of simplified CPU code.
LLM-based GPU code generation is also active: omniCUDA~\cite{gruzewski2025omnicuda} targets OpenMP-to-CUDA, KernelBench~\cite{ouyang2025kernelbench} evaluates kernel generation, and CUDA-LLM~\cite{chen2025cudallm} and CudaForge~\cite{zhang2025cudaforge} use iterative generation and hardware feedback for CUDA optimization.

LLMs have also been studied for HPC code translation, including Fortran$\to$C++ translation data and multi-turn translation~\cite{chen2024fortran2cpp,ranasinghe2025fortran}, cross-language and C-to-CUDA translation corpora~\cite{coderosetta2024,wen2022babeltower}, and Fortran$\to$C++/C++$\to$CUDA dialogue-based data generation with compilation, execution, and unit-test feedback~\cite{chen2025beyondcodepairs}.
These demonstrate LLM-based porting but do not evaluate removing CPU-specific optimizations before retranslation and reoptimization.
LLMs have been studied for general-purpose code refactoring~\cite{shirafuji2023refactoring,liu2024refactoring,cordeiro2025refactoring}: they reduce code size and cyclomatic complexity on programming-exercise submissions~\cite{shirafuji2023refactoring} and can identify and recommend refactorings in real-world projects~\cite{liu2024refactoring}, though with notable limitations~\cite{cordeiro2025refactoring}; relatedly, they exhibit a natural simplification bias even when the prompt asks for the opposite~\cite{detomasi2025simplicity}.
LLM-driven refactoring has also been applied to parallel scientific codes for energy-aware optimization~\cite{dearing2025lassi}.
Related work on LLM-based deobfuscation~\cite{beste2025deobfuscation} and input-token reduction~\cite{wang2024slimcode} further motivates simplification before translation; we evaluate LLM-based deoptimization specifically for C++$\to$CUDA porting.

\section{Common Evaluation Setting}
\label{sec:experiment}

\subsection{Task and Environment}

The task is LLM-based porting of heavily CPU-optimized C++ HPC kernels to CUDA: each input is a hand-optimized C++ implementation for aarch64/Neoverse-V2 using NEON intrinsics or other CPU-oriented optimizations, which we translate to CUDA and evaluate for feasibility and performance.

Experiments were performed on a GH200 system with a 72-core Arm Neoverse-V2 CPU and an NVIDIA H100 GPU.
The software stack is NVHPC 25.9, CUDA 12.6, cuDNN 9.5.1, NVPL 25.9, and ArmPL 25.07.1.
CPU C++ is compiled with \texttt{nvc++} (\texttt{-O3 -fopenmp -tp=neoverse-v2 -std=c++17}) and CUDA with \texttt{nvcc} (\texttt{-O3 -arch=sm\_90 -std=c++17}).

Each run uses all CPU cores with threads pinned and their memory bound to the local NUMA node.
Each candidate has one warm-up and five timed runs (minimum reported); GPU timings use CUDA events after data placement and exclude host-device transfers and validation.

\subsection{Models and Inference}

We evaluate two open-weight LLMs independently in all workflows: \texttt{gpt-oss-120b} (O120)~\cite{openai2025gptoss}, a 117B-parameter MoE (${\sim}$5.1B active) with controllable reasoning, and \texttt{qwen-3-235b-a22b-instruct-2507} (Q235)~\cite{qwen2025qwen3}, a ${\sim}$235B-parameter MoE (${\sim}$22B active) general-purpose non-thinking instruct model.
For inference, both models are served through the Cerebras inference API\footnote{\url{https://www.cerebras.ai/inference}} with temperature 0.5, \texttt{max\_completion\_tokens=16384}, and no fixed random seed (O120 uses \texttt{reasoning\_effort=medium}, the gpt-oss default); serving-side numerical precision is provider-controlled and was not recorded.

\subsection{Metrics and Statistical Analysis}

Input and generated code are normalized by removing C/C++ comments and blank lines while preserving string literals; all reported lines-of-code (LOC) values use this normalized code.

Performance differences between Direct and Deopt-Reopt are tested by a two-sided Mann-Whitney U test ($\alpha{=}0.05$) over successful trials, separately from feasibility; cases with fewer than three successes are not tested.
Unadjusted p-value marks are exploratory, while primary significance uses the Benjamini--Hochberg false discovery rate (BH-FDR) correction~\cite{benjamini1995fdr} ($\alpha{=}0.05$) separately for Single-shot and Iterative.
The Direct-3 control is auxiliary, so D+R vs Direct-3 uses unadjusted p-values only.
Feasibility is reported separately, as per-mode success rates in Table~\ref{tab:workflow_summary}.

\subsection{Benchmark Kernels}

We use the twelve HPC kernels in Table~\ref{tab:benchmarks}.
$B$ denotes batch size, $M,N,K$ principal dimensions, $C$ input channels, $H,W$ spatial sizes, $BS$ block size, and $T$ iterations.
Before observing results, we group kernels by how CPU-oriented structure relates to the natural GPU strategy: divergent-style (CPU optimizations obscure a different GPU mapping), dependency/convention-dominated (constrained by algorithmic dependences or numerical conventions), and shared-style (CPU-side structural hints still help on the GPU).
This is an author-defined interpretive framework, not a deterministic predictor.

\begin{table*}[!t]
\caption{Benchmark kernels and evaluation settings. Groups are divergent-style (Div.), dependency/convention-dominated (Dom.), and shared-style (Shr.); `*' denotes hand-coded or different-precision reference performance. Tol. is the per-element relative-error tolerance; Conv2D also accepts a $10^{-4}$ absolute error.}
\label{tab:benchmarks}
\centering
\footnotesize
\setlength{\tabcolsep}{5.5pt}
\renewcommand{\arraystretch}{1.0}
\begin{tabular}{@{}l|l|>{\raggedright\arraybackslash}p{32mm}|>{\raggedright\arraybackslash}p{50mm}|r|c|r|rr|c@{}}\hline\hline
Group & Kernel & Problem & CPU-input structure and
& LOC & Unit & Input & \multicolumn{2}{c|}{Ref. perf} & Tol. \\
 & & & GPU-side interpretation & & & perf & CPU & GPU & \\\hline
Div. & Conv2D & 2D convolution (FP32), N=16, C=32, K=64, H=W=96
& Filter-transpose cache, blocking, NEON SIMD; GPU tiles with shared memory.
& 142 & GFlops & 2328 & 474.9 & 24647 & $10^{-3}$ \\
Div. & BFFT & Batched complex FP64 FFT, B=8192, N=1024
& Radix-4 butterflies, bit reversal, twiddles; small FFTs remap differently on GPU.
& 123 & GFlops & 598.9 & 596.4 & 4907 & $10^{-8}$ \\
Div. & Softmax & Row-wise FP64 softmax, M=1024, N=32768
& Handwritten NEON \texttt{exp}, SIMD, buffering; maps to warp/block reductions.
& 67 & GB/s & 185.6 & 210.3* & 1235 & $10^{-8}$ \\
Div. & BGEMM & Batched FP64 GEMM, B=1024, M=N=K=128
& NEON FP64 microkernel, blocking, prepacking; GPU uses batch/tile parallelism.
& 210 & GFlops & 1935 & 159.4 & 32848 & $10^{-10}$ \\\hline
Dom. & DFSpMM & Sparse--dense, double-float (DF) arithmetic (48-bit significand), CSR, 32 nnz/row, M=K=24576, N=128
& Sparse access and DF arithmetic dominate; structure constrains the GPU mapping.
& 148 & GFlops & 17.3 & 735.3* & 2181* & $10^{-7}$ \\
Dom. & FFT & Single large complex FP64 FFT, N=4194304
& Radix-4 butterflies, bit reversal, twiddles, NEON wrapper; stage dependences dominate.
& 519 & GFlops & 209.8 & 170.0 & 3154 & $10^{-8}$ \\
Dom. & BTDMA & Batched FP64 tridiagonal solve, B=32768, N=256
& Parallel Thomas, two-element elimination; forward/backward dependences dominate.
& 92 & GB/s & 295.7 & 159.4 & 141.1 & $10^{-8}$ \\
Dom. & DDGEMM & Dense GEMM, double-double (DD) arithmetic (106-bit significand), M=N=K=256
& DD arithmetic dominates runtime over parallelization.
& 86 & GFlops & 44.1 & 4.3* & 569.9* & $10^{-22}$ \\\hline
Shr. & Stencil & 5-point FP64 Jacobi stencil, N=3072, T=100
& NEON SIMD, buffering, locality optimization; grid-update strategy reusable on GPUs.
& 116 & GB/s & 1404 & 970.8* & 2789* & $10^{-8}$ \\
Shr. & SpMM & Sparse--dense (FP64), CSR, 32 nnz/row, M=98304, K=524288, N=128
& p-loop unrolling, output tiling; CSR row-wise processing is a useful GPU hint.
& 98 & GB/s & 414.6 & 203.9* & 752.7 & $10^{-10}$ \\
Shr. & GEMM & Dense FP64 GEMM, M=N=K=1024
& NEON FP64 microkernel, blocking, prepacking; data-movement intent stays useful.
& 259 & GFlops & 305.5 & 2653 & 40895 & $10^{-8}$ \\
Shr. & SpMV & FP64 BSR banded sparse matrix--vector, M=K=524288, BS=4
& Block-sparse format, expansion, vectorization are useful GPU hints.
& 63 & GB/s & 1508 & 685.6 & 2527 & $10^{-10}$ \\
\hline
\end{tabular}
\end{table*}

Table~\ref{tab:benchmarks} reports LOC, input/reference performance, units, and the per-kernel validation tolerance.
Input perf is the measured performance of the hand-optimized CPU C++ input (the porting source); Ref. perf is the vendor-library or hand-coded reference on the CPU and the GPU.
These references are correctness oracles (and rough performance landmarks), not performance-tuned, same-precision competitors, so a reference can be slower than the hand-optimized input. For example, the DDGEMM reference runs in binary128 as a precision oracle, and the Conv2D reference builds an unvectorized im2col buffer before calling a library GEMM, rather than fusing the convolution like the hand-tuned input.
Outputs are validated against vendor libraries where available (NVPL/ArmPL on the CPU; cuDNN/cuFFT/cuBLAS/cuSPARSE on the GPU) and otherwise against high-precision or hand-coded references (e.g., binary128 for DDGEMM and an FP64 reference for DFSpMM).
Correctness is checked by benchmark-specific validation drivers at the benchmarking problem sizes, using per-element relative error $|x - x_{\mathrm{ref}}| / \max(|x_{\mathrm{ref}}|, 10^{-12})$ against the listed tolerance; Conv2D additionally accepts a maximum absolute error of $10^{-4}$.
DFSpMM and DDGEMM use double-word arithmetic~\cite{dekker1971}, representing a value as an unevaluated sum of two native words: double-double (DD) uses two FP64 words (a ${\sim}$106-bit significand) and double-float (DF) is the FP32-word counterpart.
DDGEMM and DFSpMM use the operation-count definitions of their GEMM and SpMM counterparts.
Except for FP32 Conv2D, interfaces are FP64.

\section{Single-shot Workflow}
\label{sec:simple_workflow}

\subsection{Method}

We first use a one-pass Single-shot workflow to observe the basic effect of LLM-based deoptimization under three modes.
\begin{itemize}
\setlength{\itemsep}{0pt}
\setlength{\parskip}{0pt}
  \item Deopt-Reopt (D+R): input code $\rightarrow$ deoptimization $\rightarrow$ CUDA translation $\rightarrow$ reoptimization, using three LLM calls.
  \item Direct (D): input code $\rightarrow$ CUDA translation $\rightarrow$ reoptimization (no deoptimization), using two LLM calls.
  \item Direct-3 (D3, control): Direct translation followed by reoptimization v1 and reoptimization v2, matching the three LLM calls of D+R; D3 trials are newly sampled and are not a subset of Direct trials. If translation does not produce a passing code (PASS), the trial stops; if v1 fails, v2 is skipped, and if v2 fails, v1 is adopted.
\end{itemize}

Direct-3 tests whether a D+R successful-trial performance advantage can be explained only by the larger number of LLM calls; it gives Direct an extra reoptimization pass and serves as an auxiliary call-count control for successful-trial performance, not a success-rate control.
Because Direct-3 spends its extra call on reoptimization rather than retrying translation, it does not test whether the same budget used for additional translation or repair attempts would close Direct's feasibility gap.
The deoptimization phase removes architecture-specific optimizations while preserving the function signature, problem sizes, and numerical results within tolerance; only code passing the same validation driver and tolerance (Table~\ref{tab:benchmarks}) as the input C++ is accepted.
The translation phase converts the (possibly deoptimized) C++ code to a complete CUDA implementation preserving the entry-point contract: the generated function takes device pointers and implements only device-side computation, while host allocation and host-device transfers are handled by the benchmark harness.
Reoptimization then improves the CUDA code for the target GPU without prescribing specific techniques.
If a required stage fails to produce a passing adopted code, the trial is counted as a failure; when a later reoptimization stage fails, the last passing CUDA code is retained.

Prompts specify the phase objective, input code, source/target languages, contract and numerical constraints, and previous-stage code or evaluation results; target hardware information is given to the translation and reoptimization prompts but not the deoptimization prompt (a hardware-agnostic step).
We run 50 independent trials per mode and model (150 per kernel-model pair, 3600 in total) and evaluate success rate and the distribution/median of final performance over successful trials; following the common evaluation setting of Section~\ref{sec:experiment}, the three modes share the LLM, temperature, prompts, problem sizes, validation, and hardware.

\subsection{Results}

\begin{figure*}[t]
\centering
\begin{minipage}[t]{0.485\textwidth}
\centering
\includegraphics[width=\linewidth]{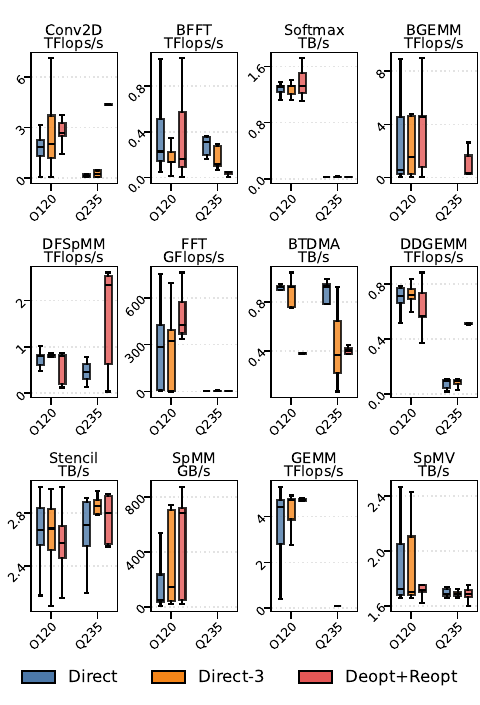}
\localfigcaption{Single-shot final performance (50 trials). D/D3/DR denote successful trials.}
\label{fig:simple_cuda_boxplots}
\end{minipage}
\hfill
\begin{minipage}[t]{0.485\textwidth}
\centering
\includegraphics[width=\linewidth]{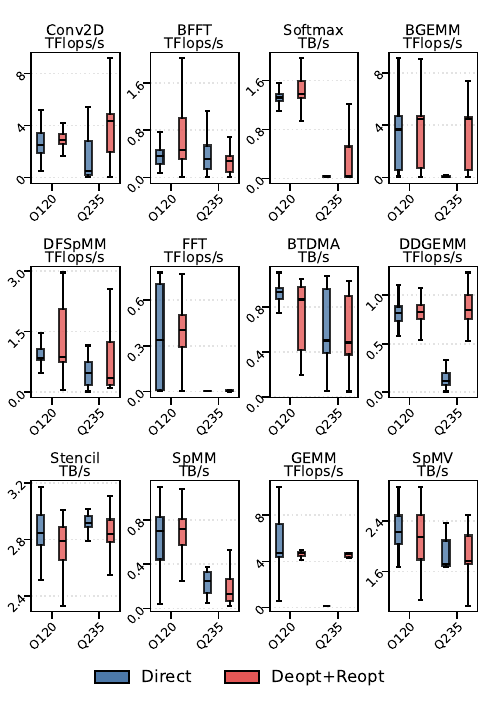}
\localfigcaption{Iterative final performance (50 trials). D/DR denote successful trials.}
\label{fig:full_cuda_boxplots}
\end{minipage}
\end{figure*}

Figure~\ref{fig:simple_cuda_boxplots} shows the Single-shot workflow results.
Both this figure and Figure~\ref{fig:full_cuda_boxplots} (Iterative) use a $3{\times}4$ kernel grid; each panel places the two models side by side with boxplots over successful trials for Direct, Direct-3 (Single-shot only), and Deopt-Reopt.

Success rates vary widely by model and kernel.
O120 obtained successful trials for all twelve kernels, whereas Q235 had 0/50 successes in both Direct and Deopt-Reopt for \texttt{spmm} and \texttt{gemm}.
At the same time, Q235 achieved near-complete success on \texttt{spmv} under Direct, with Deopt-Reopt at a high but lower rate.
Feasibility thus depends more on model--kernel compatibility than on model strength alone.
Accordingly, Table~\ref{tab:workflow_summary} should be read with success rate and successful-trial performance together, not median performance alone.
Feasibility shifts are strongly model-dependent: Deopt-Reopt sharply raises Q235 success on most kernels, consistent with Q235 being anchored to the input CPU implementation, whereas for O120 the effect is mixed and can even lower feasibility.

Across the two models, among 24 cases the Mann-Whitney U test was applied to 18 testable cases.
After BH-FDR correction within the Single-shot main series, eight cases were significant: five favored Deopt-Reopt (Q235 \texttt{conv2d} and \texttt{ddgemm}, O120 \texttt{conv2d}, \texttt{spmm}, and \texttt{gemm}) and three favored Direct (O120 \texttt{ddgemm} and \texttt{btdma}, Q235 \texttt{bfft}).
Magnitudes vary widely --- DR/D ratios from over $40\times$ down to ${\approx}0.15\times$ (Direct faster) --- so the performance effect is genuinely mixed rather than uniformly favoring Deopt-Reopt.
These cases span all three benchmark groups rather than concentrating in one: only \texttt{conv2d} favors Deopt-Reopt on both models, while \texttt{ddgemm} reverses between models (Deopt-Reopt for Q235, Direct for O120).
The benefit is thus kernel- and model-specific rather than a property of a single kernel class, and on some dependency-dominated kernels (O120 \texttt{ddgemm} and \texttt{btdma}) discarding the CPU structure instead hurts.

To check whether the D+R successful-trial performance advantage merely reflects its extra LLM call, we ran the auxiliary Direct-3 control (three LLM calls, no deoptimization; a performance, not success-rate, control) for all 24 cases with 50 trials each, of which 19 were testable.
Deopt-Reopt was significantly above Direct-3 (unadjusted) in four cases, with median DR/D3 ratios of $1.3\times$ (O120 \texttt{conv2d}), $1.3\times$ (O120 \texttt{fft}), $18\times$ (Q235 \texttt{conv2d}), and $5.4\times$ (Q235 \texttt{ddgemm}); conversely Direct-3 was faster in four others, with D3/DR ratios of $1.3\times$ (O120 \texttt{ddgemm}), $1.0\times$ (O120 \texttt{dfspmm}), $2.4\times$ (O120 \texttt{btdma}), and $2.5\times$ (Q235 \texttt{bfft}).
Of the five Single-shot main-series Deopt-Reopt advantages, three (\texttt{conv2d} on both models and Q235 \texttt{ddgemm}) remained significant against Direct-3, so part of the advantage reflects the deoptimization phase itself rather than call count alone; on the other kernels the call-count-matched control no longer favors Deopt-Reopt. Full Direct-3 statistics are in the released artifact.

Overall, Single-shot Deopt-Reopt most clearly helps \texttt{conv2d} (both models) and sharply raises Q235 feasibility, but the successful-trial performance comparison is mixed --- Direct is significantly faster on several kernels (O120 \texttt{ddgemm} and \texttt{btdma}, Q235 \texttt{bfft}) --- and Deopt-Reopt costs about 50\% more LLM calls than Direct.

\section{Iterative Workflow}
\label{sec:iterative}

\subsection{Method}

Single-shot is sensitive to one LLM output, so we also use an Iterative workflow that repeatedly plans, generates, evaluates, repairs, and selects candidates.
This compares the two modes under a search process closer to practical HPC porting, where both receive the same repair and iteration opportunities.

Each phase (deoptimization, translation, reoptimization, and direct translation) runs for up to three generations.
In each generation:
\begin{enumerate}
\setlength{\itemsep}{0pt}
\setlength{\parskip}{0pt}
\item the LLM acts as a project manager (PM) proposing a strategy from the current code and results;
\item three programmer (PG) calls generate complete implementations;
\item generated code is compiled, executed, and validated;
\item each PG may repair its code once on error; and
\item the best correct candidate seeds the next generation or the final result.
\end{enumerate}

Direct omits deoptimization and passes the input code directly to translation.
PM, PG, and repair calls use the same LLM and parameters, with role differences expressed only in the prompt.
In a single PM call, the model examines the current code, current performance, and previous-generation PG self-analyses, then emits $N$ per-PG strategies (here $N{=}3$) that each PG follows in the same generation.
Because the three PGs share one PM call, tests use the final trial-level performance value rather than treating PGs as independent samples.
Repair is triggered when a PG's code fails to compile, fails to execute, or fails validation; the failing code, the captured error message, and a brief automated self-analysis are passed back to the same PG, which produces a fixed candidate that is re-validated.

For deoptimization selection, we use LOC as a simple and interpretable proxy for simplification: it reflects removal of SIMD expansion and loop unrolling, but not portability as a whole.
Among candidates that pass validation, deoptimization adopts the smallest-LOC code subject to being smaller than the input CPU LOC, whereas reoptimization adopts the smallest-runtime code.
When no valid candidate is smaller than the input, the input is carried forward unchanged; this occurred in only 1.0\% (11/1110) of successful Iterative Deopt-Reopt trials, so the arm is essentially genuine deoptimization.
If any phase fails to obtain a valid candidate, the trial is counted as a translation failure.
Concretely, a phase fails only when none of its candidates passes validation across all generations and repair attempts (here, up to $3{\times}3{\times}2$ attempts: three generations, three PG candidates per generation, and two attempts per candidate---one initial attempt plus one repair).
Each Iterative trial is more expensive than Single-shot; we run 50 trials per kernel-mode for both models, evaluating translation success rate, execution performance, and LOC.
The twelve kernels therefore produce 2400 Iterative trials, with Deopt-Reopt's extra-phase cost included; success-rate comparisons are within-model.

\subsection{Results}

\begin{figure*}[t]
\centering
\includegraphics[width=\hsize,trim=4 8 4 2,clip]{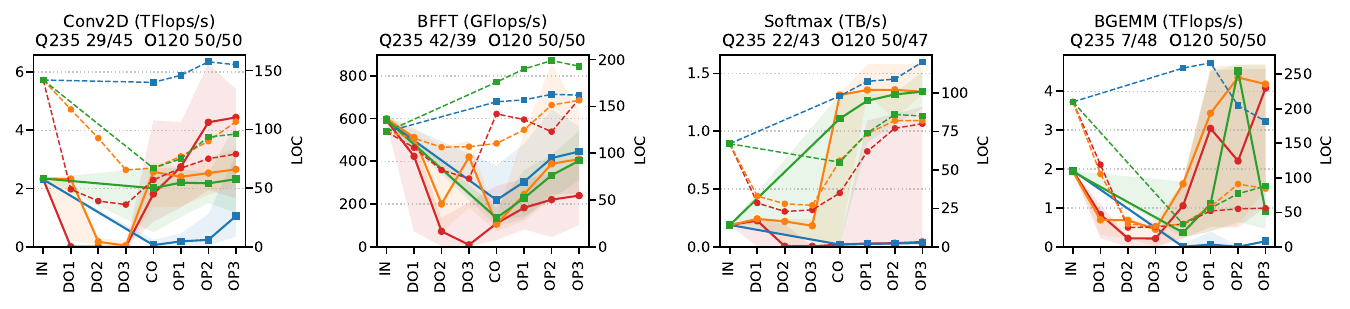}\\[1.2mm]
\includegraphics[width=\hsize,trim=4 8 4 2,clip]{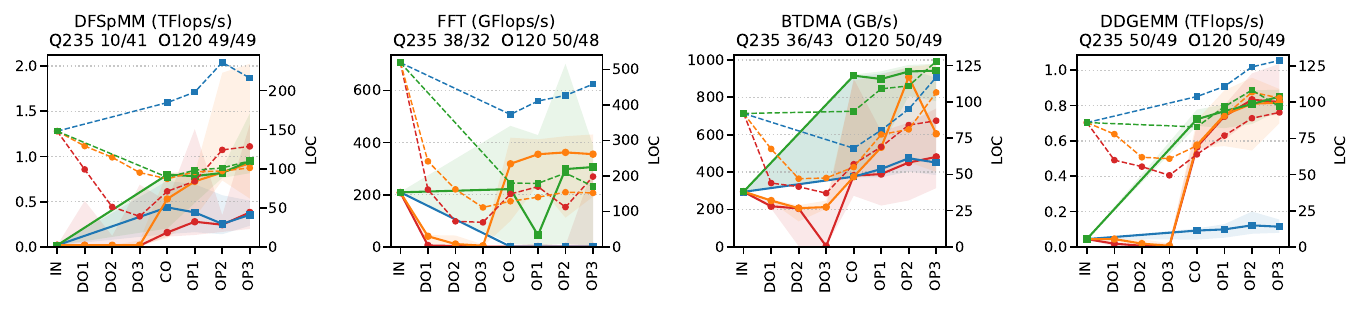}\\[1.2mm]
\includegraphics[width=\hsize,trim=4 8 4 2,clip]{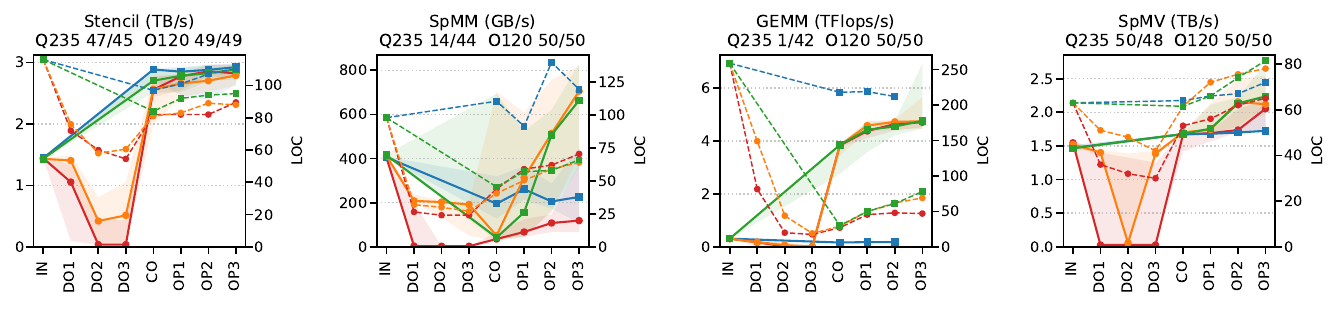}
\includegraphics[width=\hsize]{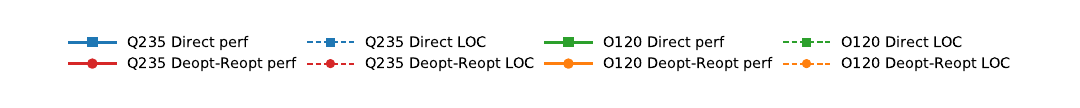}
\caption{Performance and LOC progression in Iterative (50 trials) over up to three generations. The horizontal axis traces the workflow stages: IN (input C++), DO1--DO3 (deoptimization generations), CO (CUDA translation), and OP1--OP3 (reoptimization generations); Direct has no DO stage. Solid line with band: performance median and Q1--Q3 (left axis); dashed line: LOC (right axis). Marker shape encodes mode (square: Direct, circle: Deopt-Reopt) and color encodes model. Each panel title lists per-model Direct/Deopt-Reopt success counts (out of 50 trials), consistent with Table~\ref{tab:workflow_summary}.}
\label{fig:full_cuda_progression_grid}
\end{figure*}

Results are shown in Figure~\ref{fig:full_cuda_boxplots}, with performance and LOC progression in Figure~\ref{fig:full_cuda_progression_grid}.

In Iterative, repeated generation, repair, and selection partially mitigated the Single-shot feasibility problems, which were concentrated in Q235: Deopt-Reopt recovered several Q235 kernels with few or no Single-shot successes, while O120 retained successful trials for all twelve kernels.
Of the 24 cases, 23 were testable; after BH-FDR correction within Iterative seven were significant --- five favored Deopt-Reopt (Q235 \texttt{conv2d}, \texttt{ddgemm}, and \texttt{bgemm}; O120 \texttt{bfft} and \texttt{softmax}) and two favored Direct (O120 and Q235 \texttt{stencil}).
Only the three Q235 cases carry a large effect: \texttt{conv2d} and \texttt{ddgemm} reach DR/D median ratios of $8\times$ and $7\times$, and for \texttt{bgemm} the gap is feasibility-driven (Direct succeeds in only 7/50 trials, mostly near-failures). The remaining four significant cases are far smaller --- O120 \texttt{bfft} at $1.3\times$, and O120 \texttt{softmax} and the two \texttt{stencil} cases within a few percent.
For O120, then, iteration largely closes the Single-shot gap by letting Direct also adapt toward the GPU.
For Q235 \texttt{fft}, both workflows have median performance near 1 GFlops; the outputs pass validation but stay far below the GPU reference, i.e., they do not reach high performance.
A large median gap need not be significant: several sizeable median differences in Table~\ref{tab:workflow_summary} are not statistically significant, so medians must be read together with the test.

Zero-success Single-shot recovery is most visible for Q235, where PM-driven strategy updates and one allowed repair provide a recovery path.
Deopt-Reopt markedly raises Q235 Iterative feasibility, recovering several kernels (\texttt{bgemm}, \texttt{spmm}, \texttt{gemm}, \texttt{dfspmm}) that Direct rarely compiles.

Overall, Iterative narrows the mode difference markedly for O120, whereas Q235 retains substantial Deopt-Reopt advantages on \texttt{conv2d}, \texttt{ddgemm}, and \texttt{bgemm}.

\begin{table*}[!t]
\caption{Comparison of Direct (D) and Deopt-Reopt (DR): success rate over all trials and median performance over successful trials. A marker on a median means that mode is significantly faster among successful trials --- $^*$ unadjusted p$<$0.05, $^\dagger$ also after BH-FDR correction; $^\S$ marks a median over fewer than three successful trials (not tested). Performance units follow the Unit column in Table~\ref{tab:benchmarks}.}
\label{tab:workflow_summary}
\centering
\footnotesize
\setlength{\tabcolsep}{0.5pt}
\begin{tabular}{l|l|cc|cc|cc|cc|cc|cc|cc|cc}\hline\hline
 & & \multicolumn{8}{c|}{O120} & \multicolumn{8}{c}{Q235} \\
\multirow{3}{*}{Group} & \multirow{3}{*}{Bench}
 & \multicolumn{4}{c|}{Single-shot (50 trials)}
 & \multicolumn{4}{c|}{Iterative (50 trials)}
 & \multicolumn{4}{c|}{Single-shot (50 trials)}
 & \multicolumn{4}{c}{Iterative (50 trials)} \\
 & & \multicolumn{2}{c|}{Success (\%)} & \multicolumn{2}{c|}{Median perf}
   & \multicolumn{2}{c|}{Success (\%)} & \multicolumn{2}{c|}{Median perf}
   & \multicolumn{2}{c|}{Success (\%)} & \multicolumn{2}{c|}{Median perf}
   & \multicolumn{2}{c|}{Success (\%)} & \multicolumn{2}{c}{Median perf} \\
 & & D & DR & D & DR & D & DR & D & DR & D & DR & D & DR & D & DR & D & DR \\\hline
\multirow{4}{*}{Divergent}
 & Conv2D  & 64 & 70 & 1824 & 2704$^{*\dagger}$ & 100 & 100 & 2512 & 2908$^*$ & 12 & 96 & 106 & 4366$^{*\dagger}$ & 58 & 90 & 545 & 4389$^{*\dagger}$ \\
 & BFFT    & 48 & 52 & 228 & 164 & 100 & 100 & 370 & 467$^{*\dagger}$ & 12 & 66 & 309$^{*\dagger}$ & 46 & 84 & 78 & 319 & 273 \\
 & Softmax & 94 & 90 & 1315 & 1317 & 100 & 94 & 1322 & 1385$^{*\dagger}$ & 2 & 52 & 20$^\S$ & 18 & 44 & 86 & 35 & 34 \\
 & BGEMM   & 88 & 90 & 552 & 4546$^*$ & 100 & 100 & 3666 & 4485 & 0 & 94 & --- & 354 & 14 & 96 & 1.79 & 4474$^{*\dagger}$ \\\hline
\multirow{4}{*}{Dominated}
 & DFSpMM  & 86 & 82 & 801 & 797 & 98 & 98 & 850 & 860 & 4 & 88 & 463$^\S$ & 2346 & 20 & 82 & 485 & 355 \\
 & FFT     & 42 & 56 & 288 & 424 & 100 & 96 & 336 & 402 & 2 & 20 & 0.909$^\S$ & 0.958 & 76 & 64 & 1.08 & 0.979 \\
 & BTDMA   & 92 & 44 & 923$^{*\dagger}$ & 380 & 100 & 98 & 940$^*$ & 869 & 8 & 10 & 927 & 404 & 72 & 86 & 504 & 486 \\
 & DDGEMM  & 66 & 82 & 714$^{*\dagger}$ & 564 & 100 & 98 & 819 & 826 & 42 & 80 & 89 & 510$^{*\dagger}$ & 100 & 98 & 119 & 848$^{*\dagger}$ \\\hline
\multirow{4}{*}{Shared}
 & Stencil & 82 & 58 & 2670$^*$ & 2570 & 98 & 98 & 2849$^{*\dagger}$ & 2788 & 36 & 62 & 2708 & 2799 & 94 & 90 & 2919$^{*\dagger}$ & 2840 \\
 & SpMM    & 66 & 54 & 51 & 685$^{*\dagger}$ & 100 & 100 & 700 & 723 & 0 & 0 & --- & --- & 28 & 88 & 251 & 130 \\
 & GEMM    & 84 & 66 & 4416 & 4731$^{*\dagger}$ & 100 & 100 & 4732 & 4748 & 0 & 0 & --- & --- & 2 & 84 & 181$^\S$ & 4676 \\
 & SpMV    & 68 & 60 & 1719 & 1714 & 100 & 100 & 2216 & 2139 & 96 & 72 & 1687 & 1686 & 100 & 96 & 1711 & 1760 \\
\hline
\end{tabular}
\end{table*}

Table~\ref{tab:workflow_summary} summarizes the win/loss pattern; Single-shot and Iterative each use 50 trials.
Because these medians are conditional on success, end-to-end performance also depends on the success rate: weighting a conditional median by success rate can reverse the favored mode, so the medians should be read together with the success columns.
The three-group classification has clear exceptions --- the shared-style kernel O120 \texttt{spmm} shows a large Single-shot Deopt-Reopt advantage, the divergent-style kernel Q235 \texttt{bfft} favors Direct, and on the divergent-style O120 \texttt{conv2d} Direct catches up under Iterative --- so it is only an interpretive framework.

\section{Conclusion}
\label{sec:conclusion}

We evaluated Deopt-Reopt for LLM-based C++-to-CUDA porting on twelve HPC kernels.
In Single-shot, Deopt-Reopt was significantly faster among successful trials in five testable cases --- most clearly for \texttt{conv2d}, where the CPU optimization structure diverges from the natural GPU strategy --- but Direct was faster in three (O120 \texttt{btdma} and \texttt{ddgemm}, Q235 \texttt{bfft}), so removing CPU-specific optimizations is not universally beneficial; an exploratory Direct-3 control left the call-count-matched comparison roughly balanced.
In Iterative, repeated generation and repair narrowed the gap, markedly for O120; Q235, however, retained large Deopt-Reopt advantages on \texttt{conv2d}, \texttt{ddgemm}, and \texttt{bgemm} (after BH-FDR correction, seven of 23 testable cases stayed significant, but only these three carry a large effect).
Because performance is measured over successful trials, a lower success rate can offset a conditional gain, so the evidence supports a conditional-performance benefit, not a guaranteed end-to-end one.
In sum, Deopt-Reopt is an effective but non-universal technique for LLM-based GPU porting, with gains that depend on the kernel, the model, the search budget, and the success rate.

The study is limited to two LLMs, one GPU environment, twelve kernels, a single prompt-template family, and LOC as a proxy for simplification.
Correctness was validated at the benchmark problem sizes; generalization to unseen sizes, layouts, sparsity patterns, and boundary cases remains untested.
A fully failure-inclusive analysis of the conditional comparison (best-of-N under a fixed call budget, bootstrap intervals) is left to future work. The small per-case samples also yield low statistical power, leaving some large median gaps non-significant.
Future work includes larger codebases, more model families, automatic control of deoptimization degree, and simplification metrics beyond LOC; we report median ratios as a rough magnitude indicator, while formal effect-size estimation is deferred to future analysis.
The experimental code and data are available at \url{https://github.com/mukunoki/deopt_reopt}.

\section*{Acknowledgment}
This work was supported by the ``Joint Usage/Research Center for Interdisciplinary Large-scale Information Infrastructures (JHPCN)'' in Japan (Project ID: jh260065), JSPS KAKENHI JP25K24387, and the JST Next-Generation Edge AI Semiconductor Research and Development Project JPMJES2511.

\bibliographystyle{IEEEtran}
\bibliography{paper}

\end{document}